\documentclass{article}
\usepackage{gensymb}
\usepackage{amsmath}
\usepackage{physics}

\usepackage{bm}

\DeclareRobustCommand{\uvec}[1]{{%
  \ifcsname uvec#1\endcsname
     \csname uvec#1\endcsname
   \else
    \bm{\hat{\mathbf{#1}}}%
   \fi
}}	

\begin{document}

\null  
\nointerlineskip  
\vfill
\let\snewpage \newpage
\let\newpage \relax
\begin{center}

\textbf{\huge Methods for Finding Analytic Solutions for Time Dependent Two-Level Quantum Systems and Its Generalizations}

\vspace{1cm}
\Large Rajath Krishna R, SRFP:381

\Large Jawaharlal Nehru Centre for Advanced Scientific Research (JNCASR) \\
Bangalore, India.\\
\vspace{1cm}
\Large Guide: Prof N.S. Vidhyadhiraja, \\ Theoretical Sciences Unit, JNCASR.
\end{center}
\let \newpage \snewpage
\vfill 
\break 

\begin{center}
\textbf{\LARGE Acknowledgments}

\end{center}
\vspace{1.5cm}

\large This work would not have been possible without the guidance of Prof. N.S. Vidhyadhiraja,  and I thank him for the same. I also thank Prof. Robert J Joynt, University of Wisconsin-Madison for his valuable suggestions and comments on my work. I thank the members of the lab in which I was working during the period of my internship for their constant help and support and also the administration department of JNCASR for the facilities and hospitality provided during that time. 

\break

\begin{center}
\textbf{\LARGE Abstract}
\end{center}
\vspace{1.5cm}

\large Two-level systems are one of the most important quantum systems and they form the basis of quantum computers. We briefly look at the traditional approach to two-level systems with an external driving field as well as those subjected to noise. This project is aimed at studying two specific methods for obtaining analytic solutions for two-level systems. One of the methods enables us to obtain analytic solutions for driven time-dependent two-level systems while the other attempts to give exact solution of qubit decoherence using a transfer matrix method. A thorough study of both papers is done and results are reproduced. The latter method is generalized for a qutrit system as well as a two qubit system subjected to noise. A general method is formally derived for an N-dimensional quantum system and the difficulties in applying the method in real life systems is discussed.   

\break

\section{Introduction}
\large Two-level systems are the basic constituents of quantum computers and though they are the simplest quantum systems, studying them and its interaction with the environment gives great insights into how quantum mechanical systems behave. Except for some famous special cases like the Landau-Zener model \cite{lz}, The Rabi problem \cite{r}, Jaynes-Cummings Model \cite{jc} obtaining analytic solutions for the evolution of two-level systems are extremely difficult. Analytic solutions of such systems are important for qubit control operations, self-induced transparency and for studying decoherence. 

No real system is isolated and hence, it is important to consider the interaction of a quantum system with its environment. Two-level systems coupled to an environment were largely studied using the Caldeira-Leggett model for two-state systems \cite{cl}, also called as the Spin-Boson model. Firstly, we look at a rather new method introduced by Edwin Barnes and S. Das Sarma which provides a recipe to form Hamiltonians which is guaranteed to give analytic solutions. Secondly, we study a transfer matrix approach towards finding exact solutions for two-state systems coupled to an environment under certain approximations. We formally generalize the method for a N-dimensional quantum system with special emphasis on qutrit and two-qubit systems. Also, we discuss the limitations of this method when applied to a real life system.

\section{Driven time-dependent two level systems}
\large Two state systems which are driven by an external time-dependent field are very difficult to solve. Some of the special cases for which an analytic solution has been possible include  Landau-Zener model, The Rabi problem, Jaynes-Cummings Model etc. We will take a brief survey of these models.
\subsection{Landau-Zener Model}
Provides the probability of transition between the two quantum states coupled by an external field of constant amplitude and time dependent frequency. 
   
    \[H=\begin{bmatrix}
    \Delta &  \omega_R\\
    \omega_R & \Delta 
  \end{bmatrix}
   \]      
   \vspace{-0.02cm}
   
   $\Delta$ is the detuning parameter and $\omega_R$ is the Rabi frequency. 
   We have $\omega=\omega_o+\Delta$ and we sweep over various values of $\omega$ at a rate $\dot \omega$. 
   
\subsection{The Rabi Problem}   
The response of an two-level atom to an applied harmonic electric field like for eg: $E(t)=\vec E_0cos{\omega t}$.
    $$H=H_0+H^{'}$$ where $H^{'}=- d\cdot \vec E$, $d$ is the dipole operator. 
    
\subsection{Jaynes-Cummings Model}    
Describes a two state system interacting with a bosonic field.
   $$H=H_{field}+H_{atom}+H_{int}$$ where $$H_{field}=\hbar \omega_c \hat a^+\hat a $$
   $$H_{atom}=\hbar \omega_a \sigma_z/2$$ 
   $$H_{int}=\frac{\hbar \Omega}{2} \hat E \hat S$$
   where $\hat{a}^+$ and $\hat{a}$ are bosonic creation and annihilation operators respectively. $\hat{E}=\hat{a}+\hat{a}^+$ is the field operator and $\uvec{S}=\hat{\sigma}_=+\hat{\sigma}_-$ is the polarization operator. 
    
\subsection{A recipe for finding Hamiltonians with guaranteed analytic-solutions}
In 2012, Edwin Barnes and S. Das Sarma \cite{as} developed a completely new theoretical approach towards obtaining analytic solutions for driven two-state systems. This new method gives an unbounded set of analytically solvable driven-two level systems which are driven by a single axis field. It is shown that the driving field and the evolution operator of such a system can be obtained from a single real-function satisfying some conditions.  

The Hamiltonian in consideration has the following form: $$H=\frac{J(t)}{2}\sigma_z + \frac{h}{2}\sigma_x$$ which describes any two level system driven along a single-axis. 

Transforming to the rotating x-basis we have, $$\ket{+}(t)=e^{-iht}\ket{+} and \ket{-}(t)=e^{iht}\ket{-}$$

    Also, we have the state vector in the rotating basis, $$\ket{\Psi(t)}=d_+(t)\ket{+}(t)+d_-(t)\ket{-}(t)$$ 

    Plugging into the time-dependent Schrodinger's equation and simplifying we get, $$\dot d_{\pm}(t)=-i\frac{J(t)}{2}e^{\pm iht}d_{\mp}(t)$$

    We have the unitary operator in which the elements satisfy: 
         \[U=\begin{bmatrix}
               u_{11}  & -u_{21}^*\\
               u_{21} & u_{11}^* 
            \end{bmatrix}      \]

                We\hspace{0.10cm} have\hspace{0.10cm} the\hspace{0.10cm} unitary\hspace{0.10cm} evolution\hspace{0.10cm} (in\hspace{0.10cm} the\hspace{0.10cm} z-basis)\hspace{0.10cm} of\hspace{0.10cm} the\hspace{0.10cm} wavefunction:
            $$\ket{\Psi(t)}=U(t)\ket{\Psi(0)} =c_1(0)U(t)\ket{\uparrow}+c_2(0)U(t)\ket{\downarrow} \quad$$ 
  
Transforming to the x-basis and then transforming to the rotating x-basis we get, $$\ket{\Psi(t)}=[D_+(t)c_1(0)+D_-^*(t)c_2(0)]\ket{+}(t)  +[D_-(t)c_1(0)-D_+^*(t)c_2(0)]\ket{-}(t)$$

      $$ where \quad D_\pm=\frac{1}{\sqrt{2}}e^{\pm iht/2}(u_{11}\pm u_{21})$$

    Comparing the two equations obtained for $\ket{\Psi(t)}$ we get,
      $$\dot D_\pm=-i\frac{J(t)}{2}e^{\pm iht}D_\mp$$

 The two coupled first-order differential equations can be combined to give one second order differential equation.    

$$\ddot D_+ + (-ih-\dot J/J)\dot D_++(J^2/4)=0$$
\\

A reverse engineering approach is adopted and the differential equation is solved for $J(t)$ instead of $D_+$ which gives: 
$$J(t)=\pm \frac{\dot D_+e^{-iht}}{\sqrt{c-\frac{1}{4}D_+^2e^{-2iht}-\frac{ih}{2}\int_{0}^{t} dt e^{-2iht}D_+^2(t)}}$$and back substitution gives: $$D_-=\pm 2i \sqrt{c-\frac{1}{4}D_+^2e^{-2iht}-\frac{ih}{2}\int_{0}^{t}e^{-2iht}D_+^2(t) dt}$$

We take the following ansatz which preserved unitarity $$D_+=e^{i(F-K+ht)}cos(\Phi)$$
    $$D_-=e^{-iK}sin(\Phi)$$  and the following relations are obtained:
    $$sin(2\Phi)=sec(F)e^{h\int_{0}^{t} tan(F)}$$ and, $$J(t)=2\dot Ksec(F)tan(\Phi)$$
    
We will take the initial condition to be $D_+(0)=D_-(0)=\frac{1}{\sqrt{2}}$ which in turn gives $\Phi(0)=\frac{\pi}{4}$, $F(0)=K(0)=0.$    
  \\  
Taking F to be of the form $F=arctan(\frac{\dot q}{hq})$ gives $$sin(2\Phi)=\sqrt{q^2+\frac{\dot q^2}{h^2}}$$ and 
      $$J=\frac{\ddot q+ h^2q}{\sqrt{h^2(1-q^2)-\dot q^2}}$$
      
The initial conditions on $F$, $K$ and $\Phi$ translate to $$q(0)=1, \quad \dot q(0)=0 \quad \ddot q(0)=-h^2$$

Also, we have the constraint on the function q given by, $$\dot q^2 \leq h^2(1-q^2)$$
   
Any function satisfying the above initial conditions and constraint is guaranteed to produce an analytical solution for the evolution of the system.        

\section{Two-level system with noise}
Decoherence problems are usually solved by coupling the system concerned with an environment. Then, a master equation for the reduced density matrix is formed which can take care of the effect of the system on the environment too. A successful model which implements the above method is called the Spin-Boson model. 

\subsection{Spin-Boson Model}
It describes a quantum particle in one dimension coupled to a bath of infinite harmonic oscillators. The Hamiltonian for this model has the following form \cite{w} : 
$$H=H_0+H_B+H_{int}$$
where $H_0$ is the free Hamiltonian of the spin-1/2 system. $$H_B=\sum_{k} \frac{p_k^2}{2m_k}+\frac{1}{2}m_k\omega_k^2x_k^2$$ and $$H_{int}=\sigma_z \sum_{k} \lambda_k x_k$$. Here, $p_k$, $m_k$, $\omega_k$ and $x_k$ are the momentum, mass, frequency and position respectively of the $k^{th}$ harmonic oscillator.

$\lambda_k$ is the strength of coupling between oscillator and spin. The system in consideration eventually loses coherence due to this coupling. 

Now, a master equation can be formulated for the reduced density matrix. This, method can also take into account the "back-action" of the system on the environment. 

\subsection{A transfer matrix approach}
In certain cases, this back-action is not important and we can model environment as a source of noise. A transfer matrix method \cite{tm} can be used to obtain exact solution for such systems.
\\

 We will consider the following form of the Hamiltonian:
 $$H=-B_0 \sigma_z-\vec b(t)\cdot \vec \sigma$$
 were $\vec b(t)$ is the random function.
 We will work in the Heisenberg piture and thus, we  will be concerned with the evolution of the operators $\sigma_x$, $\sigma_y$ and $\sigma_z$ which together with identity forms a complete operator basis for a two-state quantum system. 
 
 We will assume that the the function $\vec b(t)$ is piecewise constant with discontinuous jumps at regular intervals of length $\tau$. $$\vec B(t)=B_0\uvec{z}+\vec b_1=\vec B_1 \quad \quad 0< t < \tau$$ 
 $$\vec B(t)=B_0\uvec{z}+\vec b_2=\vec B_2 \quad \quad \tau< t < 2\tau \quad etc.$$
 \\
 Define $H_i=-\vec B_i \cdot \vec \sigma=-\vec B_0\sigma_z-\vec b_i \cdot \vec \sigma $
 and $$U_i=e^{-iH_i\tau}$$
 
 We will also assume that there is no correlation between the different $\vec b_i$ and that each $\vec b_i$ has the same probability distribution $P(\vec b)$.
 
 The expectation value of the operators at time $\tau$ is given by: 
 $$\overline{\vec \sigma(\tau)=U_i^+\vec \sigma(0)U_i}$$
 
Simplifying the above expression:
 
 \[ \overline{\left[ E \cos(B_1\tau) - i {\hat B_1}\cdot{\vec \sigma}\sin(B_1\tau)\right]{\vec \sigma}(0) \left[ E \cos(B_1\tau) + i {\hat B_1}\cdot{\vec \sigma}\sin(B_1\tau)\right]} = \\ \]
 
  \[= \overline{\cos^2(B_1\tau)} \;{\vec \sigma}(0)  + i {\vec \sigma}(0) \left[\overline{\cos(B_1\tau)\sin(B_1\tau){\hat B_1} } \cdot{\vec \sigma}\right] - \] \\ 
 
 $$- i \left[ \overline{\cos(B_1\tau)\sin(B_1\tau){\hat B_1} } \cdot{\vec \sigma} \right] {\vec \sigma}(0) + \overline{\sin^2(B_1\tau)\left[{\hat B_1} \cdot{\vec \sigma}\right] {\vec \sigma}(0) \left[{\hat B_1}  \cdot{\vec \sigma} \right] }=\\$$

$$= \overline{\cos^2(B_1\tau)} \;{\vec \sigma}(0) + i \sum_i{\left(\overline{\cos(B_1\tau)\sin(B_1\tau)B_{1,i} } \right) \left[ {\vec \sigma}(0)\sigma_i - \sigma_i  {\vec \sigma}(0)\right]} +$$
 $$ \quad \quad \quad \quad \quad \quad \hspace{6cm} \sum_{i,j}{\left( \overline{\sin^2(B_1\tau)B_{1,i} B_{1,j}  } \right) \sigma_i \;{\vec \sigma}(0)\; \sigma_j}$$
 
 Which we can write as: 
 $$\vec \sigma(\tau)=I_0\vec \sigma(0)+ \sum_{i} I_i[\vec \sigma(0)\sigma_i-\sigma_i\vec \sigma(0)] +\sum_{ij} I_{ij}[\sigma_i\vec \sigma(0)\sigma_j] $$ where,
 
 $$I_0=\int P(\vec b)cos^2(B\tau)d^3b $$
 $$I_i=\int P(\vec b)\uvec{B}_i sin(B\tau)cos(B\tau)d^3b$$ 
 $$I_{ij}=\int P(\vec b) \uvec{B}_i \uvec{B}_j  sin^2(B\tau)d^3b$$
 
 Expanding the above summation and writing the coefficients of the Pauli matrices as a matrix gives:
  $$\left[ \begin{array}{c} \overline{ \sigma_x(\tau)} \\ \overline{ \sigma_y(\tau)} \\ \overline{ \sigma_z(\tau)} \end{array} \right] = \begin{bmatrix} T_{xx} & T_{xy} & T_{xz} \\ T_{yx} & T_{yy} & T_{yz} \\ T_{zx} & T_{zy} & T_{zz} \end{bmatrix} \times \left[ \begin{array}{c} \overline{\vec \sigma_x(0)} \\ \overline{ \sigma_y(0)} \\ \overline{ \sigma_z(0)} \end{array} \right]$$
  or
  $$\overline{ \vec \sigma(\tau)}=T\vec \sigma(0)$$
  
  where   $$ T_{xx}=I_0+I_{xx}-I_{yy}-I_{zz} \quad  T_{yy}=I_0-I_{xx}+I_{yy}-I_{zz}$$ $$T_{zz}=I_0-I_{xx}-I_{yy}+I_{zz} $$
  $$T_{xy}=2I_{xy}+2I_z ,\quad T_{yx}=2I_{xy}-2I_z$$ 
  $$ T_{xz}=2I_{xz}-2I_y ,\quad T_{zx}=2I_{xz}+2I_y $$
  $$T_{yz}=2I_{yz}+2I_x ,\quad T_{xy}=2I_{yz}-2I_x $$
  
  After m time steps we have: 
  $$\overline{ \vec \sigma(m\tau)}=T^m\vec \sigma(0)$$
  
  The transfer matrix can be diagonalized and hence, $T^m$ can be easily calculated giving exact solutions for the expectation value of the operators. 
  
  \section{An attempt to extend the transfer matrix method to other systems}
  
  \subsection{Three State(Qutrit) System with Noise}
 We will choose the 8 Gell-Mann matrices as the operator basis. 
Consider the following Hamiltonian for a three state system: $$H=-\vec B\cdot \vec \lambda.$$
were $\vec B=B_0 \uvec{i} +\vec b$ where $\vec b$ is the random vector with components as the three random functions which will get coupled to the three Gell-Mann matrices $\vec\lambda=(\lambda_1,\lambda_2,\lambda_2)$.

Assume that the random vector is piecewise constant with discontinuous jumps at regular time intervals of length $\tau$. Therefore, we have $$B_i=B_0 \uvec{i} +\vec b_i$$ for the $i^{th}$ time slot. 
We will also assume that there is no correlation between the different $\vec b_i$ and that each $\vec b_i$ has the same probability distribution $P(\vec b)$.

Also, we have the unitary operator given by $U_i=e^{\frac{-iH_i\tau}{\hbar}}$ where $H_i=\vec B_i\cdot \vec \lambda$. 

Since, the Gell-Mann matrices $\lambda_1$,$\lambda_2$ and $\lambda_3$ form a SU(2) subgroup of $SU(3)$ they anti-commute. Defining $\beta=\tau|\vec B|/\hbar$.

$$U_1=\sum_{n=0}^{\infty} \frac{(i)^n\beta^n(\hat{B}\cdot \vec \lambda)^n}{n!}$$ 
$$=\sum_{n=0}^{\infty} \frac{(-1)^n\beta^{2n}(\hat{B}\cdot \vec \lambda)^{2n}}{(2n)!}+i\sum_{n=0}^{\infty} \frac{(-1)^n\beta^{2n+1}(\hat{B}\cdot \vec \lambda)^{2n+1}}{(2n+1)!}$$

Now,
$$(\hat{B}\cdot \vec \lambda)^2=(\sum_{i}^{3} B_i\lambda_i)^2=\sum_{ij}^{3} B_iB_j\lambda_i\lambda_j$$

Since, $\lambda_1,\lambda_2,\lambda_3$ anti-commute with each other the above expression reduces to \[(\hat{B}\cdot \vec \lambda)^2=\sum_{i}^{3}B_i^2A=A \]

Substituting and simplifying we get,

\begin{equation}
 U_1=I+i\beta (\hat{B}\cdot \vec \lambda)+A[cos(\beta)-1]+iA(\vec B\cdot \vec \lambda)[sin(\beta)-\beta] 
\end{equation}

where $\beta=\frac{\tau |\vec B|}{\hbar}$ and  \[
A=
  \begin{bmatrix}
   1 0 0\\
   0 1 0\\
   0 0 0\\
     
  \end{bmatrix}
\]
\\
\\

Proceeding in a way analogous to that given in the previous section to obtain the transfer matrix results in a 9x9 transfer matrix instead of a 8x8 one. This, is because unlike the Pauli Matrices the Gell-Mann matrices are not closed under multiplication(apart form a multiplicative constant). Thus, while deriving the equations for the expectation values of the operators we will have to invoke the identity matrix making the number of operators considered 9 instead of the 8 Gell-Mann matrices alone.  
\vspace{0.5cm}

A more transparent method which also removes the difficulty in dealing with identity is obtained by considering the evolution of the density matrix itself and its decomposition in terms of the operator basis. 
The most general form of the density matrix after a time $t$ can be written as \cite{bq}:
$$\rho (t)=\frac{1}{3}I+\sum_{a=1}^{8} \rho_{a}(t)\lambda_{a}$$

The $\rho_{\alpha}$ are real numbers which completely characterize the state of the system. 

Also, density matrix evolved in a unitary fashion in the following way as $\rho(\tau)=\overline{U_1\rho(0)U_1^+}$
So, we can write: $$\rho (\tau)=\overline{U_1\Big [\frac{1}{3}I+\sum_{b=1}^{8} \rho_{b}(0)\lambda_{b}\Big ]U_1^+}$$
$$==\frac{1}{3}I+\overline{U_1 \Big [\sum_{b=1}^{8} \rho_{b}(0)\lambda_{b}\Big ]U_1^+}$$
Then we have,  $$\sum_{a=1}^{8} \rho_{a}(\tau)\lambda_{a}=\overline{U_1 \Big [\sum_{b=1}^{8} \rho_{b}(0)\lambda_{b}\Big ]U_1^+}$$

We know that for the Gell-Mann matrices $Tr(\lambda_a\lambda_b)=2\delta_{ab}$. Thus, we can simplify the above expression by multiplying it with $\lambda_c$ and taking the trace.Then we get,
$$\rho_{c}(\tau)=\frac{1}{2}Tr\Big (\lambda_c \overline{U_1 \Big [\sum_{b=1}^{8} \rho_{b}(0)\lambda_{b}\Big ]U_1^+} \Big )$$ which we can write as, $$\rho_{c}(\tau)=\sum_{b=1}^{8} T_{cb}\rho_b(0)$$
where $$T_{cb}=\frac{1}{2}Tr ( \lambda_c \overline{U_1\lambda_b U_1^+})$$
$$=\frac{1}{2}Tr\Big(\hspace{0.1cm}\overline{[I+i\beta (\hat{B}\cdot \vec \lambda)+A[cos(\beta)-1]+iA(\vec B\cdot \vec \lambda)[sin(\beta)-\beta]]\lambda_b}$$ 

$$\hspace{5cm}\overline{[I-i\beta (\hat{B}\cdot \vec \lambda)+A[cos(\beta)-1]-iA(\vec B\cdot \vec \lambda)[sin(\beta)-\beta]]}\lambda_c \Big )$$

\subsection{Two-Qubit System with Noise}
The observables in the Hilbert space corresponding to a two-qubit system can be spanned by the following set of 16 3x3  matrices.
$$\{\sigma_i \otimes \sigma_j: 0 \leq i,j \leq 3 \}$$

where $\sigma_1,\sigma_2,\sigma_3$ are the Pauli matrices and $\sigma_0=I$. Here $\otimes$ denotes the Kronecker product. Note that the subsets $$\{\sigma_i \otimes \sigma_0: 0 \leq i \leq 3 \},
\quad 
\{\sigma_0 \otimes \sigma_j: 0 \leq j \leq 3 \}$$
acts as Pauli matrices. 

Hence, we can write down a Hamiltonian in which the coupling to the noise function is through any of the six matrices in the above subset. This will enable us to apply the same ideas as in section 3.2. 
\\
Consider the Hamiltonian,
$$H=-\vec B\cdot \vec e.$$
where $\vec B=B_0 \uvec{i} +\vec b$ where $\vec b$ is the stochastic vector with components as the three stochastic functions which will get coupled to the Gell-Mann matrices $\vec e=(e_4,e_8,e_{12})$. 
We will assume that the random functions and the probability distribution has properties already assumed in the previous two sections.

Since the matrices $e_4,e_8,e_{12}$ have properties of Pauli matrices we can derive the following identity for the exponentiation of the Hamiltonian $U_i=e^{\frac{-iH_i\tau}{\hbar}}$:

$$U_1=Icos(\beta)+i(\hat{B}\cdot \vec e)sin(\beta)$$ 

where  $\beta=\frac{\tau |\vec B|}{\hbar}$.
\vspace{0.5cm}

We can consider the general form of a 4x4 density matrix and obtain the transfer matrix through the method followed in the previous section. 

For two-qubit system we obtain the following equation for the coefficients of the density matrix:
$$\rho_{c}(\tau)=\sum_{b=1}^{15} T_{cb}\rho_b(0)$$ where

$$T_{bc}=\frac{1}{4}Tr ( e_c \overline{U_1e_b U_1^+})$$
$$=\frac{1}{4}Tr\Big(\hspace{0.1cm}\overline{[Icos(\beta)+i(\vec B\cdot \vec e)sin(\beta)]e_b[Acos(\beta)-iA(\vec B\cdot \vec e)sin(\beta)]}e_c \Big )$$

\subsection{N-dimensional Quantum System with Noise}

The transformations of a N-dimensional quantum system belongs to SU(N) and the algebra is an $N^2-1$ dimensional space. The transfer matrix will be $N^2-1$x$N^2-1$. 
A suitable orthogonal basis is the set of Generalized Gell-Mann matrices \cite{bq} $\lambda_i$ where $i=1,2,,...N^2-1$. Expanding a general NxN density matrix in terms of the $\lambda_i$s we get:
$$\rho_{c}(\tau)=\frac{1}{2}Tr\Big (\lambda_c \overline{U_1 \Big [\sum_{b=1}^{N^2-1} \rho_{b}(t)\lambda_{b}\Big ]U_1^+} \Big )$$ which we can write as, $$\rho_{c}(\tau)=\sum_{b=1}^{N^2-1} T_{cb}\rho_b(0)$$
where $$T_{cb}=\frac{1}{2}Tr ( \lambda_c \overline{U_1\lambda_b U_1^+})$$

Now, if \( \lambda_{k_1},..,\lambda_{k_n}\) form some sub-algebra of $SU(N)$ such that $\lambda_{k_i}$ anti-commute with each other and squares to a matrix A which scales to A, we can use these matrices to model the Hamiltonian for the system. Then, as before we can derive the identity: 
$$U_1=I+i\beta (\hat{B}\cdot \vec \lambda)+A[cos(\beta)-1]+iA(\vec B\cdot \vec \lambda)[sin(\beta)-\beta]$$  where $\vec \lambda=(\lambda_{k_{1}},...,\lambda_{k_{n}})$ and  $\beta=\frac{\tau |\vec B|}{\hbar}$.
\vspace{0.5cm}

Using the above identity we can write the transfer matrix as:
$$T_{cb}=\frac{1}{2}i\sum_i \Big[Tr(\overline{\beta(cos(\beta)-1)\hat{B_i}}[\lambda_c\lambda_i\lambda_bA-\lambda_cA\lambda_b\lambda_i])+$$
$$Tr(\overline{(sin(\beta)-\beta)(cos(\beta)-1)\hat{B_i}}[\lambda_cA\lambda_i\lambda_bA-\lambda_cA\lambda_bA\lambda_i])+$$
$$Tr(\overline{(sin(\beta)-\beta)\hat{B_i}}[\lambda_cA\lambda_i\lambda_b])-Tr(\overline{\beta\hat{B_i}}[\lambda_c\lambda_b\lambda_i-\lambda_c\lambda_i\lambda_b])\Big]+$$ 
$$\frac{1}{2}\sum_{ij} \Big[Tr(\overline{\beta(sin(\beta)-\beta)\hat{B_i}\hat{B_j}}[\lambda_c\lambda_i\lambda_bA\lambda_j-\lambda_cA\lambda_i\lambda_b\lambda_i])+ Tr(\overline{(sin(\beta)-\beta)^2\hat{B_i}\hat{B_j}}[\lambda_cA\lambda_iA\lambda_j])+$$
$$ Tr(\overline{\hat{B_i}\hat{B_j}\beta^2}[\lambda_c\lambda_i\lambda_b\lambda_j])\Big]+\frac{1}{2}\overline{(cos(\beta)-1)^2}[\lambda_cA\lambda_bA]+\frac{1}{2}\overline{cos(\beta)-1}[\lambda_cA\lambda_b+\lambda_c\lambda_bA]+\frac{1}{2}\lambda_c\lambda_b$$
\vspace{0.5cm}

If the system in consideration is a N-qubit system then, we can take the operator basis to be: 
$$\{\sigma_{i_{1}} \otimes ... \otimes \sigma_{i_{N}}: 0 \leq i_1,i_2,...i_N \leq 3 \}$$

where $\sigma_{1_{j}},\sigma_{2_{j}},\sigma_{3_{j}}$ are the Pauli matrices and $\sigma_{0_{j}}=I$ where j=1,..,N.
Block diagonal elements within this basis forms different subsets such that the elements in teach of the subsets have properties as that of Pauli matrices. Any, one such subset can be used to model the Hamiltonian. Then, as before we can derive the  identity:

$$U_1=Icos(\beta)+i(\hat{B}\cdot \vec e)sin(\beta)$$ 

where  $\beta=\frac{\tau |\vec B|}{\hbar}$.
\\
Using the above identity we can write the transfer matrix as:
$$T_{cb}=\frac{1}{c}\Big[ Tr(\overline{cos(\beta)}e_ce_b)+i\sum_i Tr( \overline{sin(\beta) cos(\beta) \hat{B_i}}[e_ce_ie_b-e_ce_be_i])+$$
$$\hspace{7cm}\sum_{ij} Tr(\overline{sin^2(\beta) \hat{B_i}\hat{B_j}}[e_ce_ie_be_j])\Big]$$
where c is the constant in the orthogonality condition. 

In general, simplifying the transfer matrix to obtain exact solutions is not trivial. 
But if the transfer matrix can be diagonalized we can form the diagonal matrix $D=diag(d_1,...,d_{N^2-1})$ where $$D=RTR^{-1}$$. Then, at time $t_f=m\tau$ steps the coefficients in the expansion of the density matrix is given by: $$\rho_i(t_f)=R^{-1}_{ij}(d_j)^{t_f/\tau}R_{jk}\rho_k(0)$$ where $i=1,...,N^2-1$

\section{Conclusion}
The method introduced by  Barnes and Das Sarma is a highly efficient one. There has been extensions of this method \cite{ash} \cite{ta} for other systems with different forms of coupling  but the idea remains the same. It is clear from equation that the function $q(t)$ should be a smooth one with at least first and second derivatives. So, it is obvious that noise functions discussed in section 3 doesn't come under this category as the simplest of them have discontinuous jumps. 

The transfer matrix method enable us to find analytic solutions for two-state systems in which the interaction with the environment can be assumed to be of a special kind i.e, where we can assume it to be a noise function which is piecewise-constant at equal intervals. The method cannot be as such applied to higher dimensional systems. A significant property of SU(2) which enabled the derivation of a transfer matrix is that apart from the operator basis(the three Pauli matrices plus identity) being closed under multiplication the set of Pauli matrices themselves have closure under multiplication. This is a crucial property which neither the 8-Gell-Mann matrices nor the 15 matrices(except identity) of the operator basis of two-qubit space have. Though, we can formally write an extended matrix in both the cases it is not easy to see how to simplify them. 

Rather, we could derive the transfer matrix for an N-dimensional quantum system by looking at the density matrix evolution. Whether this transfer matrix can be simplified to obtain exact solutions depends on the specific system under consideration. There may be situations in which the matrix is diagonalizable in which case the exact solution is easily obtained.


\begin{thebibliography}{9}
\bibitem{lz} 
 L. Landau, Phys. Z. Sowjetunion 2, 46 (1932)

\bibitem{r} 
I. Rabi, Phys. Rev. 51, 652 (1937)

\bibitem{jc}
E. T. Jaynes and F. W. Cummings, Proc. IEEE 51, 89
(1963)

\bibitem{cl}
A. J. Leggett, S. Chakravarty, A. T. Dorsey, Matthew P. A. Fisher, Anupam Garg, and W. Zwerger,  Rev. Mod. Phys. 67, 725 (1995)

\bibitem{as} 
Edwin Barnes, S. Das Sarma,
PhysRevLett.109.060401 (2012)

\bibitem{tm}
Diu Nghiem, Robert Joynt,
PhysRevA.73.032333 (2006)

\bibitem{w}
U. Weiss:
Quantum Dissipative Systems, World Scientific, 1999.

\bibitem{bq}
Reinhold A. Bertlmann, Philipp Krammer,
J. Phys. A: Math.Theor. 41 (2008) 235303

\bibitem{ash}
A Messina, H Nakazato
Journal of Physics A Mathematical and Theoretical 47(44):445302 · October 2014

\bibitem{ta}
Edwin Barnes
Phys. Rev. A 88, 013818


\end{thebibliography}
\end{document}